\documentclass
[english,aps,preprint,superscriptaddress,nofootinbib,pra]{revtex4-2}%
\usepackage[T1]{fontenc}
\usepackage[latin9]{inputenc}
\usepackage{amsmath}
\usepackage{amssymb}
\usepackage{babel}
\usepackage{amsfonts}
\usepackage{graphicx}
\usepackage{MnSymbol}

\begin{document}
\title{Microcausality and Tunneling Times in
	Relativistic Quantum Field Theory}
\author{M.\ Alkhateeb}
\affiliation{Research Unit Lasers and Spectroscopies (UR-LLS), naXys \& NISM, University of Namur, Rue de Bruxelles 61, B-5000 Namur, Belgium}
\affiliation{Centre for Mathematical Sciences, University of Plymouth, Plymouth, PL4 8AA, UK}

\author{A.\ Matzkin}
\affiliation{Laboratoire de Physique Th\'eorique et Mod\'elisation, CNRS Unit\'e 8089, CY Cergy Paris Universit\'e, 95302 Cergy-Pontoise cedex, France}

\begin{abstract}
We show, in the framework of a space-time resolved relativistic quantum field
theory approach to tunneling, that microcausality precludes superluminal
tunneling dynamics. More specifically  in this work dealing
with Dirac and Klein-Gordon fields, we first prove that microcausality holds
for such fields in the presence of a background
potential. We then use this result to show that an intervention performed on
a localized region of an initial wave packet subsequently scattering on a
potential barrier does not result in any effect
outside the light cone emanating from that region. We illustrate these results
with numerical computations for Dirac fermions and Klein-Gordon bosons.\ 

\end{abstract}
\maketitle

\newpage



\section{Introduction}

While tunneling is ubiquitous in about any field described by quantum theory,
the mechanism accounting for quantum tunneling has remained controversial \cite{soko-recent}. In particular the issue of the time it takes a \textquotedblleft
particle\textquotedblright\ to tunnel across a potential barrier is not
well-defined within quantum theory, given that on the one hand a quantum
system is not a classical particle localized on a well-defined worldline, and
on the other hand there is no time operator in quantum mechanics, and hence no
time eigenvalues that would provide an unambiguous answer. Several theoretical
approaches have been developed to capture the traversal time \cite{timeQM}, many of which
predict the possibility of superluminal times, including in first quantized relativistic
contexts
\cite{grobe-super,janner,winful,deleo,bernardini,gasparian,deleo2013,pollak2020,galapon}%
. On the other hand works addressing the issue of tunneling times within a relativistic 
quantum field
theory (QFT) framework are scarce \cite{charis}.

In a recent work \cite{causal-tunneling-PRA2025}, it was shown by employing
a space-time resolved formulation (see \cite{grobe-su-review,AMpra22} and
 \cite{lv,cqft-opt,newst1,newst11,newst2} for related work)  of  relativistic QFT that the tunneling dynamics does not exhibit any superluminal effects. That was done for the case of a Dirac field describing an electron wave packet. 
However, in that work the proof of causality relied on having an
initial wave packet defined on a compact support and launched towards a
potential barrier; causality was then addressed by modifying the initial
wave packet and proving that such a change had no effect outside the light cone
emanating from the compact support, including on any density transmitted
through the potential.

It is well-known, however, that a single-particle QFT state cannot be defined on
a compact spatial support -- it must have infinite tails. And working with a
wave packet defined over a compact support prompted us to employ in our previous work
\cite{causal-tunneling-PRA2025} a non-standard form of QFT (introduced in
\cite{compact}) based on a number symmetry rather than the fundamentally
correct charge symmetry; in particular the field operators introduced in Ref.
\cite{compact} and used to prove causality of the tunneling process in Ref.
\cite{causal-tunneling-PRA2025} are not the usual ones.

The aim of the present work is to give a proof of the causality of the tunneling
process for standard QFT states and field operators. The protocol is
different from the one elaborated for compact states, as rather than relying
on the light-cone emanating form a well-defined spatial support, we will be
led to introduce an intervention modifying a wave packet (intrinsically defined
with infinite tails) over a compact spatial support. We moreover extend the
formulation of the proof so as to cover a spin-0 bosonic field in addition to
the fermionic Dirac field. To this end, we will briefly recall in
Sec.\ \ref{sec-framework} the space-time resolved QFT formalism we will employ and how
a wave packet is propagated. In Sec.\ \ref{micro-sec}, we will introduce a protocol
that will be used to prove that wave packet tunneling dynamics respects
causality.\ We will first show that microcausality holds in the
presence of a background field. We will then use this result to prove that if two initial wave packets 
differ only within a region $\mathcal{D}$ of their spatial density,
then the density in a region causally disconnected from $\mathcal{D}$ 
on the other side of the potential barrier is identical for both wave packets.
The proof will be illustrated by working out numerical
computations in Sec.\ \ref{sec4}. We will close with a Discussion and
Conclusion section.

\section{Space-time resolved QFT with a background potential}

\label{sec-framework}

\subsection{Field operator \label{subsec-f}}

Our approach is based on a computational QFT framework \cite{grobe-su-review},
recently extended to treat particle scattering across a finite barrier
\cite{AMpra22} (see also \cite{lv,cqft-opt,newst1,newst11,newst2} for related recent work). The
formalism is essentially the same for Klein-Gordon (KG) or Dirac fields. The field
operator takes the usual form%
\begin{equation}
\hat{\Phi}(t,x)=\int dp\left(  \hat{b}_{p}(t)v_{p}(x)+\hat{d}_{p}^{\dagger
}(t)w_{p}(x)\right)  \label{phit}%
\end{equation}
where $v_{p}(x)$ and $w_{p}(x)$ are respectively the positive and negative
solutions of the first quantized free Dirac or KG equation; hence $v_{p}$ and $w_{p}$ obey%
\begin{equation}
i\hbar\partial_{t}v_{p}=H_{0}v_{p}%
\end{equation}
where $H_{0}$ is the Dirac or Klein-Gordon Hamiltonian (in the latter case
given in the so-called Feschbach-Villars form). We will consider in this work
only one spatial dimension (hence $H_{0}$ is two-dimensional both in the Dirac
case, since spin-flip does not occur, and in the KG case). The creation and
annihilation operators obey the commutation relations%
\begin{equation}%
\begin{split}
\lbrack\hat{b}_{p}^{\dagger},\hat{b}_{p^{\prime}}]_{\epsilon}  &  =[\hat{d}%
_{p}^{\dagger},\hat{d}_{p^{\prime}}]_{\epsilon}=\delta(p-p^{\prime}),\\
\lbrack\hat{b}_{p}^{\dagger},\hat{d}_{p^{\prime}}]_{\epsilon}  &  =[\hat{d}%
_{p}^{\dagger},\hat{b}_{p^{\prime}}]_{\epsilon}=0,
\end{split}
\label{creanncomm}%
\end{equation}
where $\epsilon=1$ for fermions and $\epsilon=-1$ for bosons; $\Vert0\rrangle$ defines
the vacuum state, i.e. $b_{p}\Vert0\rrangle=d_{p}\Vert0\rrangle=0$.

The full first quantized Hamiltonian is%
\begin{equation}
H=H_{0}+V(x) \label{fullH}%
\end{equation}
where $V(x)$ is a rectangular-like potential barrier. The Hamiltonian $H$
generates a unitary (or pseudo-unitary in the KG case) evolution. The QFT
Hamiltonian density is given as usual by%
\begin{equation}
\hat{\mathcal{H}}=\hat{\Phi}^{\dagger}(t,x)H\hat{\Phi}(t,x).
\end{equation}

As is well-known, it can be shown that the Heisenberg equation in the Dirac or
KG cases becomes%
\begin{equation}
i\partial_{t}\hat{\Phi}(t,x)=[\hat{\Phi}(t,x),\hat{\mathcal{H}}]=H\hat{\Phi
}(t,x)
\end{equation}
so that the time evolution of the field operator is obtained from the first
quantized unitary $U(t,t_{0})\equiv e^{-i\hat{H}(t-t_{0})}$ as%
\begin{equation}
\hat{\Phi}(t,x)=e^{-i\hat{H}t}\Phi(0,x)=U(t)\hat{\Phi}(0,x)
\end{equation}
where the operator $\hat{H}$ is applied to $w_{p}(x)$ and $v_{p}(x)$. Note
that $U$ accounts for the evolution due to the background field. Comparing the
time-evolved field operator to Eq. (\ref{phit}), one obtains the expressions
of the time-evolved creation and annihilation operators%
\begin{equation}%
\begin{split}
\hat{b}_{p}(t)  &  =\int dp^{\prime}\left(  U_{v_{p}v_{p^{\prime}}%
}(t)b_{p^{\prime}}+U_{v_{p}w_{p^{\prime}}}(t)d_{p^{\prime}}^{\dagger}\right)
\\
\hat{d}_{p}^{\dagger}(t)  &  =\int dp^{\prime}\left(  U_{w_{p}v_{p^{\prime}}%
}(t)b_{p^{\prime}}+U_{w_{p}w_{p^{\prime}}}(t)d_{p^{\prime}}^{\dagger}\right)
\end{split}
\end{equation}
where $U_{w_{p}v_{p^{\prime}}}(t)=\int dxv
w_{p}^\dagger(x)\sigma U(t)v_{p^{\prime}}(x)$; here $\sigma$ denotes the identity matrix in the case of fermions and the $\sigma_3$ Pauli matrix in the case of bosons.  
\subsection{Densities and wave packets}

The density operators for positive and negative energy states are given
by the usual expressions
\begin{equation}%
\begin{split}
\hat{\rho}_{+}(t) &  =\iint dpdp^{\prime}\hat{b}_{p}^{\dagger
}(t)\hat{b}_{p^{\prime}}(t)v_{p}^{\dagger}\sigma v_{p^{\prime}}\\
\hat{\rho}_{-}(t) &  =\epsilon\iint dpdp^{\prime}\hat{d}_{p}^{\dagger}%
(t)\hat{d}_{p^{\prime}}(t)w_{p^{\prime}}^{\dagger}\sigma w_{p}%
\end{split}
\end{equation}
where for fermions $\sigma=I$ (identity matrix) and $\epsilon=1$ while for KG
bosons $\sigma=\sigma_{3}$ is a Pauli matrix) and $\epsilon=-1$. The
charge density operator is given by%
\begin{equation}
\hat{\rho}(t,x)=\hat{\Phi}^{\dagger}(t,x)\sigma\hat{\Phi}(t,x)
\label{densdef}
\end{equation}
which can be rewritten as%
\begin{equation}%
\begin{split}
\hat{\rho}(t,x) &  =\iint dpdp^{\prime}\hat{b}_{p}^{\dagger}(t)\hat
{b}_{p^{\prime}}(t)v_{p}^{\dagger}(x)\sigma v_{p^{\prime}}(x)\\
&  -\iint dpdp^{\prime}\hat{d}_{p^{\prime}}(t)\hat{d}_{p}(t)w_{p}^{\dagger
}(x)\sigma w_{p^{\prime}}(x)\\
&  =\hat{\rho}_{+}(t,x)-\hat{\rho}_{-}(t,x)
\end{split}
\label{dendef}%
\end{equation}
where we used the anticommuting (commuting) property of the antiparticle creation
and annihilation operators in the case of fermions (bosons).

In vacuum the space-time resolved density is the expectation value $\rho_{0}(t,x)=\llangle0\Vert
\hat{\rho}(t,x)\Vert0\rrangle$.\ However we are interested in wave packet
tunneling whereby a particle is initially prepared in a state $\Vert
\chi\rrangle$. This is a single particle state (here an electron, or a
scalar boson) that can be generically written in case the particle has positive charge as
\begin{equation}
\Vert\chi\rrangle=\int dpg_{+}(p;x_{0},p_{0})b_{p}^{\dagger}(0)\Vert0\rrangle,
\label{cs1}%
\end{equation}
where $g_{+}(p)$ are the wave packet amplitudes in momentum space and $x_{0}$
($p_{0}$) denotes the initial average position (momentum). The time-dependent
density in the presence of a wave packet is given by
\begin{equation}
\rho(t,x)=\llangle\chi\Vert\hat{\rho}(t,x)\Vert\chi\rrangle. \label{tdd}%
\end{equation}
$\rho(t,x)$ accounts for the full charge density -- the part due to the
wave packet as well as the particle-antiparticle pairs created by the
potential. Note that at $t=0$ the density cannot be bounded on a compact
support (ie, it has tails spreading to infinity), a consequence of requiring a
single particle wave packet (hence defined over an expansion containing only contributions
from the positive energy sector).

\subsection{Microcausality}

Microcausality is the assertion that observables that are space-like separated
commute.\ While it is frequently considered as an axiom in some versions of
QFT \cite{aqft}, microcausality can be explicitly proved for some free quantum fields. In particular the proof that a non-interacting free KG or
Dirac field obeys microcausality is a well-known textbook result
\cite{greiner,pad-book}: if $\hat{O}(t,x)$ and $\hat{O}^{\prime}(t^{\prime
},x^{\prime})$ are two obervables then
\begin{equation}
\lbrack\hat{O}^{\prime}(t^{\prime},x^{\prime}),\hat{O}(t,x)]=0 \label{microc}%
\end{equation}
for$\;c^{2}\left(  t^{\prime}-t\right)  ^{2}-\left(  x^{\prime}-x\right)
^{2}<0$. The standard proof involves writing an arbitrary observable as a
bilinear combination of field operators,
\begin{equation}
\hat{O}(t,x)=\hat{\Phi}^{\dagger}(t,x)o(t,x)\hat{\Phi}(t,x) \label{obser}%
\end{equation}
where $o(t,x)$ is a matrix consisting of $c$-numbers \cite{greiner,pad-book}.
The commutator in Eq. (\ref{microc}) is then written in terms of the
anti-commutators (for the Dirac case) or commutators (for the KG case)
$[\hat{\Phi}^{\dagger}(t^{\prime},x^{\prime}),\hat{\Phi}(t,x)]_{\pm}$.\ For
free Dirac or KG fields, these anti-commutators can be computed explicitly in
closed form \cite{greiner} and are proved to vanish for space-like separated intervals.

Note that the density operator given by Eq. (\ref{densdef}) is the simplest
bilinear form involving field operators; this is the only observable we will
be interested in in this work.

\section{Microcausality and the impossibility of superluminal
tunneling\label{micro-sec}}

\subsection{Microcausality with a background field}

In the presence of a background field, it is generally impossible to compute
the field commutators in the case of bosons (or anti-commutators in the case of fermions) $[\hat{\Phi}^{\dagger}(t^{\prime},x^{\prime}),\hat{\Phi
}(t,x)]_{\pm}$ in closed form.\ However it is straightforward to establish
that for space--like separated points, such commutators (or anti-commutators) must vanish. This can
be seen by noting that if $x$ and $x^{\prime}$ are space-like separated, then
there is a reference frame for which $x$ and $x^{\prime}$ lie in a
hypersurface of simultaneity. In this reference frame the commutator for the
density becomes $\left[  \hat{\rho}(t,x),\hat{\rho}(t,y)\right]  $ (with
$x\neq y$) which can be readily computed as%
\begin{equation}%
\begin{split}
\left[  \hat{\rho}(t,x),\hat{\rho}(t,y)\right]   &  =\hat{\Phi}^{\dagger
}(t,x)\left(  \left[  \hat{\Phi}(t,x),\hat{\Phi}^{\dagger}(t,y)\right]
_{\epsilon}\hat{\Phi}(t,y)+\hat{\Phi}^{\dagger}(t,y)\left[  \hat{\Phi}(t,x),\hat
{\Phi}(t,y)\right]  _{\epsilon}\right) \\
&  +\left(  \left[  \hat{\Phi}^{\dagger}(t,x),\hat{\Phi}^{\dagger
}(t,y)\right]  _{\epsilon}\hat{\Phi}(t,y)+\hat{\Phi}^{\dagger}(t,y)\left[  \hat{\Phi
}^{\dagger}(t,x),\hat{\Phi}(t,y)\right]  _{\epsilon}\right)  \hat{\Phi}(t,x)\\
&  =\hat{\Phi}^{\dagger}(t,x)\left[  \hat{\Phi}(t,x),\hat{\Phi}^{\dagger
}(t,y)\right]  _{\epsilon}\Phi(t,y)+\hat{\Phi}^{\dagger}(t,y)\left[  \hat{\Phi
}^{\dagger}(t,x),\hat{\Phi}(t,y)\right]  _{\epsilon}\hat{\Phi}(t,x).
\end{split}
\label{microrho}%
\end{equation}
Since $\left[  \hat{\Phi}(t,x),\hat{\Phi}(t,y)\right]  _{\epsilon}=0$ due
to bosons (fermions) statistics we only need to compute the equal-time commutation (anti-commutation) relation
\begin{equation}
\left[  \hat{\Phi}^{\dagger}(t,x),\hat{\Phi}(t,y)\right]  _{\epsilon}=\left[  \int
dp\left(  \hat{b}_{p}^{\dagger}(t)v_{p}^{\dagger}(x)+\hat{d}_{p}%
(t)w_{p}(x)^{\dagger}\right)  ,\int dp\left(  \hat{b}_{p}(t)v_{p}(x)+\hat
{d}_{p}^{\dagger}(t)w_{p}(x)\right)  \right]  _{\epsilon} \label{commut-d}%
\end{equation}
in the presence of an electromagnetic background field. 

This involves the commutators (or anti-commutators) of the type 
\begin{equation}%
\begin{split}
&  [\hat{b}_{p_{1}}^{\dagger}(t),b_{p_{2}}(t)]_{\epsilon}=\\
&  \left[  \int dp_{1}^{\prime}\left(  U_{v_{p_{1}}v_{p_{1}^{\prime}}}^{\ast
}\hat{b}_{p_{1}^{\prime}}^{\dagger}+U_{v_{p_{1}}w_{p_{1}^{\prime}}}^{\ast}%
\hat{d}_{p_{1}^{\prime}}\right)  ,\int dp_{2}^{\prime}\left(  U_{v_{p_{2}%
}v_{p_{2}^{\prime}}}\hat{b}_{p_{2}^{\prime}}+U_{v_{p_{2}}w_{p_{2}^{\prime}}%
}\hat{d}_{p_{2}^{\prime}}^{\dagger}\right)  \right]  _{\epsilon}.
\end{split}
\end{equation}
Using Eq. (\ref{creanncomm}), one obtains%

\begin{equation}%
\begin{aligned}
&  \left[  \hat{b}_{p_{1}}^{\dagger}(t),b_{p_{2}}(t)\right]  _{\epsilon}\\
&  =\int dp_{1}^{\prime}\left(  U_{v_{p_{1}}v_{p_{1}^{\prime}}}^{\ast
}U_{v_{p_{2}}v_{p_{1}^{\prime}}} \epsilon U_{v_{p_{1}}w_{p_{1}^{\prime}}}^{\ast
}U_{v_{p_{2}}w_{p_{1}^{\prime}}}\right) \\
&  =\int dp_{1}^{\prime}\left(  \langle v_{p_{2}}|\hat{U}|v_{p_{1}^{\prime}%
}\rangle\langle v_{p_{1}^{\prime}}|\hat{U}^{\dagger}|v_{p_{1}}\rangle
\epsilon\langle v_{p_{2}}|\hat{U}|w_{p_{1}^{\prime}}\rangle \langle
w_{p_{1}^{\prime}}|\hat{U}^{\dagger}|v_{p_{1}}\rangle\right) \\
&  =\langle v_{p_{2}}|\hat{U}\hat{U}^{\dagger}|v_{p_{1}}\rangle=\langle
v_{p_{2}}|v_{p_{1}}\rangle=\delta(p_{1}-p_{2}),
=\end{aligned}
\end{equation}
where in the last line, we used the completeness relation:$\int dp^{\prime
}\left(  |v_{p^{\prime}}\rangle\langle v_{p^{\prime}}|+\epsilon|w_{p^{\prime}%
}\rangle\langle w_{p^{\prime}}|\right)  =1$, with $\epsilon=1$ in the case of
fermions and $\epsilon=-1$ in the case of bosons, and the orthonormality of
the solutions of the free Dirac or KG equations. Similarly, one can show
that:
\begin{equation}
\left[  \hat{d}_{p_{1}}^{\dagger}(t),d_{p_{2}}(t)\right]  _{\epsilon}=\delta
(p_{1}-p_{2}).
\end{equation}
Inserting these commutators (anti-commutators) into Eq. (\ref{commut-d}) leads
to%
\begin{equation}
\left[  \hat{\Phi}^{\dagger}(t,x),\hat{\Phi}(t,y)\right]  _{\epsilon}=\int
dp(e^{ip(y-x)}+e^{ip(x-y)})=\delta(x-y).
\end{equation}

Pluging-in this result into Eq. (\ref{microrho}) leads to\ $\left[  \hat{\rho
}(t,x),\hat{\rho}(t,y)\right]  =0$ for $x\neq y$ which ensures $\left[
\hat{\rho}(t,x),\hat{\rho}(t^{\prime},x^{\prime})\right]  =0$ for space-like
separated points by Lorentz transforming back to the original frame. We have
therefore shown that microcausality holds for density operators in the
presence of a background field.

\subsection{Microcausality and wave packet tunneling}

\begin{figure}[tb]
	\includegraphics[width=0.7\linewidth]{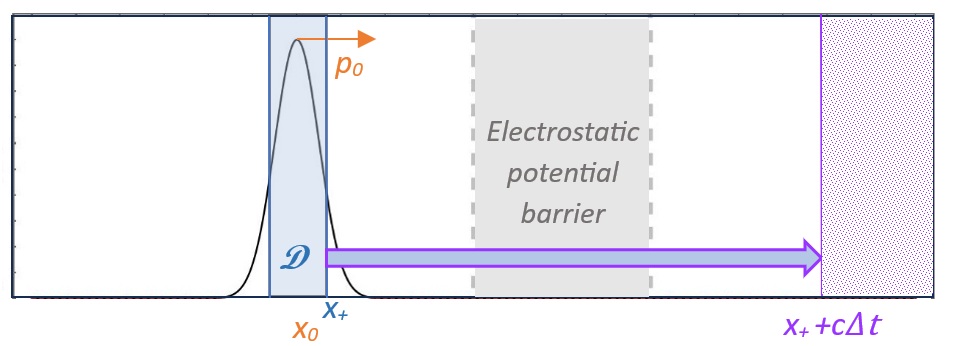}
\caption{The initial wave packet at $t_0=0$, with average position $x_0$ and momentum $p_0$ is shown
	along with the intervention region $\mathcal{D}$ defined on a compact support. The causally
	disconnected region at time $t^\prime = \Delta t - t_0$ on the other side of the potential barrier, displayed with purple dots, is defined as the region lying outside the light-cone emanating from the right edge $x_+$ of $\mathcal{D}$.
}%
	\label{fig-schema}%
\end{figure}

In order to connect microcausality with conditions on the wave packet dynamics,
consider the following situation (see Fig. \ref{fig-schema}). We start with a one-particle state
$\Vert\chi\rrangle$ given by Eq. (\ref{cs1}).\ In configuration space, this is
a wave packet with mean position $x_{0}$ and momentum $p_{0}$. As recalled
above, the density $\rho(t=0,x)=\llangle\chi\Vert\hat{\rho}(t=0,x)\Vert
\chi\rrangle$ has infinite tails.

From Eq. (\ref{cs1}), we can write the first quantized initial wave packet
amplitude $\chi(0,x)$ as \cite{schweber}
\begin{equation}
\hat{\Phi}(0,x)\Vert\chi\rrangle=\chi(0,x)\Vert0\rrangle,\label{1qwp}%
\end{equation}
and similarly
\begin{equation}
\llangle\chi\Vert\hat{\Phi}^{\dagger}(0,x)=\llangle0\Vert\chi^{\dagger}(0,x).
\end{equation}
Let us introduce a localized intervention on the initial wave packet localized
on a compact support $\mathcal{D}$.\ For definiteness, we will take this
intervention to reshape the initial wave packet with the constraint that the
total density remains constant. This can be represented by the operator%
\begin{equation}
\hat{O}(0,\mathcal{D})=1+\int_{\mathcal{D}}dxf(x)\hat{\Phi}^{\dagger}%
(0,x)\hat{\Phi}(0,x),\label{mutilation}%
\end{equation}
where $f(x)$ is a real function modifying the spatial profile of the initial
wave packet. Since the total density on $\mathcal{D}$ is conserved, we have%
\begin{equation}
\int_{\mathcal{D}}dxf(x)
\chi(x)^\dagger \sigma \chi(x)=0.\label{condi}%
\end{equation}
Note $\hat{O}$ is bilinear in $\hat{\Phi}$.

We next introduce $\hat{O}^{\prime}(t^{\prime},x^{\prime})$ as being an
observable at a spacetime point $(t^{\prime},x^{\prime})$ lying to the right
of the potential barrier and that is spacelike separated relative to any point
$(t,x)$ of $\mathcal{D}$. We are interested in the correlation function
\begin{equation}
\mathcal{C}(t^{\prime},x^{\prime};0,x)=\llangle\chi\Vert\hat{O}^{\prime
}(t^{\prime},x^{\prime})\hat{O}(0,\mathcal{D})\rrangle \label{corr}%
\end{equation}
involving the joint operations \textquotedblleft intervention on $\mathcal{D}$
and the application of some obervable $\hat{O}^{\prime}$ at $(t^{\prime
},x^{\prime})$\textquotedblright. To be specific let us take $\hat{O}^{\prime
}(t^{\prime},x^{\prime})$ to be the density $\hat{\rho}(t^{\prime},x^{\prime
})$. Using Eq.
(\ref{mutilation}) the correlation function becomes
\begin{equation}%
\begin{split}
\mathcal{C}(t^{\prime},x^{\prime};0,x)  &  =\llangle\chi\Vert\hat{\rho
}(t^{\prime},x^{\prime})\Vert\chi\rrangle+\llangle0\Vert\hat{\rho}(t^{\prime
},x^{\prime})\Vert0\rrangle\int_{\mathcal{D}}dxf(x)\chi^{\dagger}%
(0,x)\chi(0,x)\\
&  =\llangle\chi\Vert\hat{\rho}(t^{\prime},x^{\prime})\Vert\chi\rrangle.
\end{split}
\label{cause}%
\end{equation}
where we used Eq. (\ref{condi}). Hence the correlation function is simply the
density at $(t^{\prime},x^{\prime})$ and does not in any way depend on the
intervention carried out at a spacelike interval involving the initial
wave packet. If superluminal transmission were possible, one would expect the
result of the intervention to be reflected in the transmitted wave packet, in
which case Eq. (\ref{cause}) would not hold. We therefore conclude that
microcausality prevents superluminal transmission.

As a corollary, consider two initial wave packets $\Vert\chi\rrangle$ and
$\Vert\tilde{\chi}\rrangle$ that differ inside $\mathcal{D}$ but are identical
elsewhere. There is therefore an operator $\hat{O}(0,\mathcal{D})$ defined by
Eq. (\ref{mutilation}) such that $\Vert\tilde{\chi}\rrangle=\hat
{O}(0,\mathcal{D})\Vert\chi\rrangle.$ We then have
\begin{equation}
\llangle\tilde{\chi}\Vert\hat{\rho}(t^{\prime},x^{\prime})\Vert\tilde{\chi
}\rrangle=\llangle\tilde{\chi}\Vert\hat{\rho}(t^{\prime},x^{\prime})\hat
{O}(0,\mathcal{D})\Vert\chi\rrangle,\label{zequ1}%
\end{equation}
and using the commutativity of $\hat{O}(0,\mathcal{D})$ and $\hat{\rho
}(t^{\prime},x^{\prime})$ (as per microcausality) as well as the Hermiticity of $\hat{O}$, the
right-handside term becomes $\llangle\chi\Vert\hat{\rho}(t^{\prime},x^{\prime
})\Vert\chi\rrangle$ leading to
\begin{equation}
\llangle\tilde{\chi}\Vert\hat{\rho}(t^{\prime},x^{\prime})\Vert\tilde{\chi
}\rrangle=\llangle\chi\Vert\hat{\rho}(t^{\prime},x^{\prime})\Vert
\chi\rrangle.\label{zequ2}%
\end{equation}
This means that for an arbitrary intervention in the region $\mathcal{D}$ at
time $t=0$, the expectation value of the charge density $\rho(t^{\prime
},x^{\prime})$ at a space-like distant point from the intervention does not
change: the density $\rho(t^{\prime},x^{\prime})$ is identical for any
wave packets that initially differ only inside $\mathcal{D}$. Note that $\rho(t^{\prime},x^{\prime})$ 
does not vanish due to the particle-antiparticle pairs produced by the barrier, as well as 
to the dynamics of the tails of the wave packets.

\section{Illustrations}
\label{sec4}
\subsection{Setting}

In order to illustrate our main results given in the preceding Section within
our space-time resolved QFT framework, we undertake here computations of Eq.
(\ref{zequ2}).\ We specifically start with a Gaussian wave packet of mean
position and momentum $x_{0}$ and $p_{0}$ expressed in Fock space as%
\begin{equation}
\lVert\chi\rrangle=\int dpg_{+}(p)b_{p}^{\dagger}\lVert0\rrangle\label{gaup}%
\end{equation}
with $g_{+}(p)=\exp\left(  -(p-p_{0})^{2}\,\sigma^{2}\right)  \cdot\exp\left(
-ix_{0}p\right)  ,$ where $\sigma$ is the spatial width.\ The spatial
amplitude of this Gaussian wave packet at $t=0$, introduced in Eq.
(\ref{1qwp}), is denoted here $\chi(x)$. The modified density $\tilde{\chi
}(x)$ is obtained by introducing an intervention $F(x)$ on the Gaussian wave
packet such that%
\begin{equation}
\tilde{\chi}(x)=F(x)\chi(x).
\end{equation}
$F(x)$ is nonzero only within the compact support $\mathcal{D}$ and is related
to the function $f(x)$ of Eq. (\ref{mutilation}) defining the intervention $O$
by%
\begin{equation}
F(x)=\sqrt{1+f(x)}.
\end{equation}

Note that the QFT state corresponding to the wave packet $\tilde{\chi}(x)$ can
be expressed similarly to Eq (\ref{gaup}) as%
\begin{equation}
\Vert\tilde{\chi}\rrangle=\int dp\tilde{g}_{+}(p)\hat{b}^{\dagger}%
\Vert0\rrangle.
\end{equation}
The amplitudes $\tilde{g}_{+}(p)$ are obtained by Fourier transforming
$\tilde{\chi}(x)$. For definiteness we will choose the modified wave packet to
result by implementing the intervention
\begin{equation}
f(x)=\sin^{13}(x)\theta\left(  x+\frac{D}{2}\right)  \theta\left(  x-\frac
{D}{2}\right)  \label{mutif}%
\end{equation}
over $\mathcal{D}$ ($D$ is hence the width of $\mathcal{D}$). We will consider
a rectangular-like potential barrier given by
\begin{equation}
V(x)=\frac{1}{2}V_{0}\left(  \tanh\left(  \frac{x+\frac{d}{2}}{\kappa
}\right)  -\tanh\left(  \frac{x-\frac{d}{2}}{\kappa}\right)  \right)
,\label{poti}%
\end{equation}
where $V_0$ is the barrier height, $d$ is the barrier width, $\kappa$ is the smoothness parameter.

In order to compute the densities given by the expectation values
$\llangle\chi\Vert\hat{\rho}(t^{\prime},x^{\prime})\Vert\chi\rrangle$ and
$\llangle\tilde{\chi}\Vert\hat{\rho}(t^{\prime},x^{\prime})\Vert\tilde{\chi
}\rrangle$ we start from Eqs. (\ref{densdef})-(\ref{dendef}) and (\ref{tdd}) in order to
obtain the positive and negative charge densities, $\rho_{+}$ and $\rho_{-}$,
and from there the total density $\rho$. The only difference when computing
the densities for each of the two initial wave packets lies in the amplitudes, $g_{+}(p)$ and $\tilde{g}_{+}(p)$
repectively. Setting $G(p)$ to indicate either case, we compute the positive
charge density from%
\begin{equation}
\rho_{+}(t,x)=\llangle0\lVert\int dpG^{\ast}(p)\hat{b}_{p}\int dp\hat{b}%
_{p}^{\dagger}(t)v_{p}^{\dagger}(x) \sigma \int dp\hat{b}_{p}(t)v_{p}(x)\int
dpG(p) \hat{b}_p^\dagger\rVert0\rrangle
\end{equation}
which after some algebra gives%
\begin{equation}%
\begin{split}
\rho_{+}(t,x) &  =\int dp_{1}\dots dp_{3}U_{v_{p_{1}}w_{p_{2}}}^{\ast}(t)v_{p_{1}}^{\dagger
}(x)\sigma U_{v_{p_{3}}w_{p_{2}}}(t)v_{p_{3}}(x)\\
&  +\int dp_{1}dp_{2}G^{\ast}(p_{1})U_{v_{p_{2}}v_{p_{1}}}%
^{\ast}(t)v_{p_{2}}^{\dagger}(x)\sigma\int dp_{1}dp_{2}U_{v_{p_{2}}v_{p_{1}}%
}(t)v_{p_{2}}(x)G(p_{1});
\end{split}
\label{rhoP}%
\end{equation}
the unitary evolution operator $U$ and the free basis of the Klein-Gordon or
Dirac equations have been defined in Sec. \ref{subsec-f}. Similarly, the
density of anti-fermions or anti-bosons is%
\begin{equation}%
\begin{split}
\rho_{-}(x) &  =\epsilon \int dp_{1}\dots dp_{3}U_{w_{p_{1}}v_{p_{2}}}^{\ast}(t)w_{p_{1}%
}^{\dagger}(x)\sigma U_{w_{p_{3}}v_{p_{2}}}(t)w_{p_{3}}(x)\\
&  -\int dp_{1}dp_{2}U_{w_{p_{1}}v_{p_{2}}}^{\ast}(t)w_{p_{1}}^{\dagger
}(x)G^{\ast}(p_{2})\sigma\int dp_{1}dp_{3}G(p_{1})U_{v_{p_{1}}w_{p_{1}}}%
(t)w_{p_{2}}(x).
\end{split}
\label{rhoM}%
\end{equation}
In Eqs. (\ref{rhoP}) and (\ref{rhoM}), the first term corresponds to the
density of created fermions or bosons while the second term corresponds to the
time evolution of the wave packet. 

\subsection{Numerical Results}

\begin{figure}
	\includegraphics[scale=0.15]{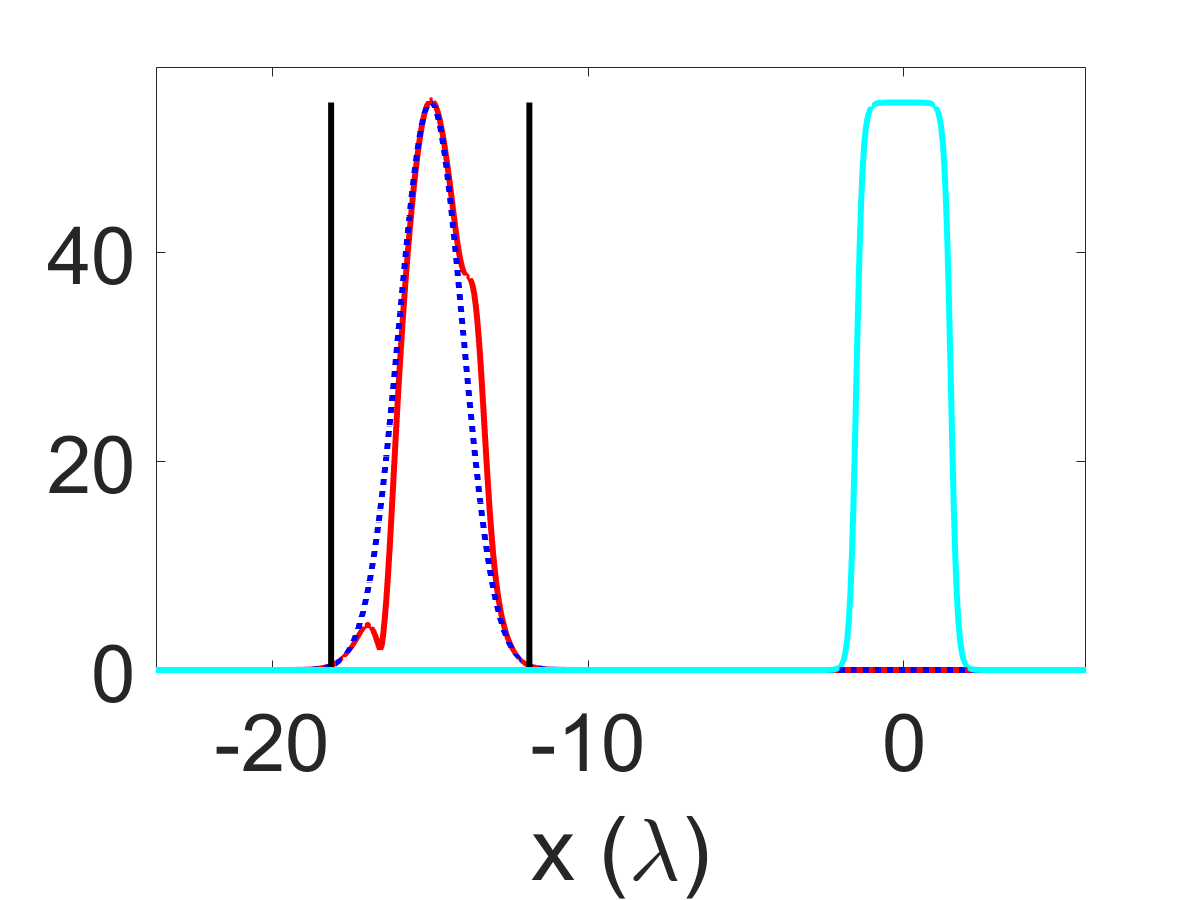}
	\includegraphics[scale=0.15]{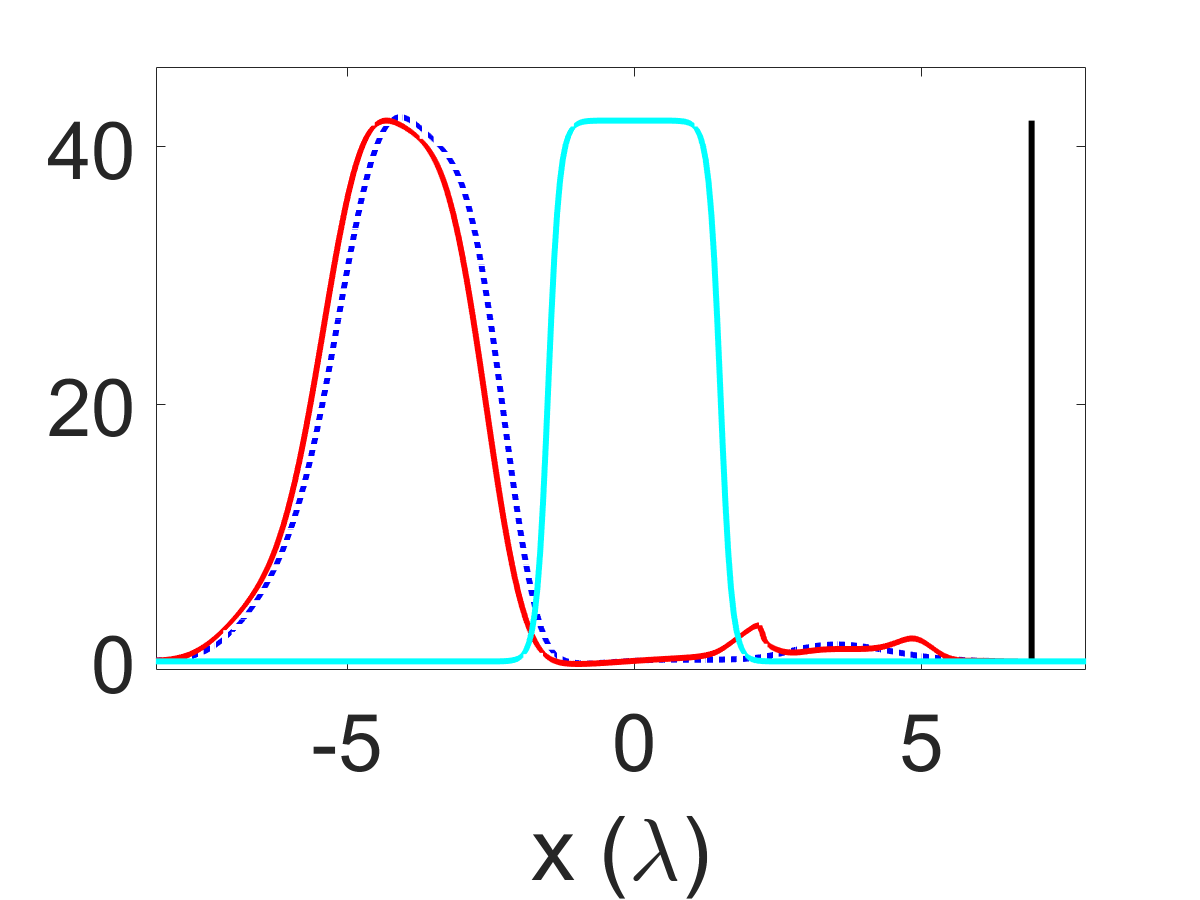}
	\includegraphics[scale=0.15]{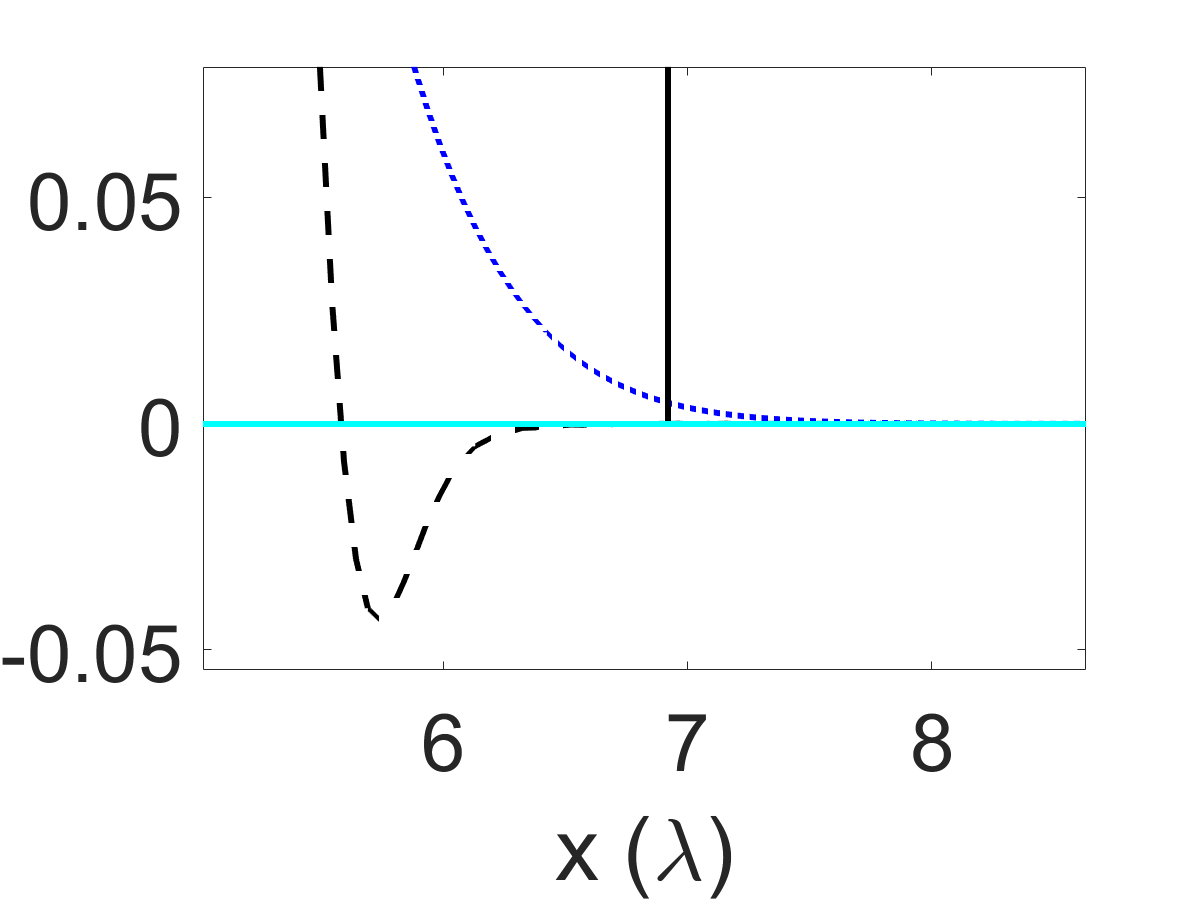}
\caption{The left panel shows, in dotted blue, an initial Dirac Gaussian wave packet of width $\sigma = 1 \lambda$ ($\lambda = \frac{\hbar}{m_e c}$ is the Compton wavelength of the electron), initial position $x_0 = -15 \lambda$, and average momentum $p_0 = \sqrt{V_0^2 - m^2 c^4}/c$, where $V_0$ is the potential barrier height. The mutilated wave packet is shown in solid red, with the intervention function defined by Eq.~(\ref{mutif}) and $D = 1 \lambda$. The vertical black lines mark the edges of the intervention area in the left panel and the light cone emanating from them in the middle and right panels. The potential barrier is given by Eq.~(\ref{poti}) with $V_0 = 2.5 mc^2$, $\kappa = 0.2 \lambda$, and $d = 3 \lambda$. The middle panel shows the reflected and tunneled wave packets at $t = 0.137 \lambda/c$. In the right panel, we zoom in on the region around the light cone, showing the tunneled part of the Gaussian wave packet in dotted blue and the difference between the Gaussian and the mutilated wave packets in dashed black. The calculations are performed on a lattice of width $100 \lambda$, with $2^{11}$ sites and time steps of $\delta t = 137 \times 10^{-6} \lambda/c$.}
\label{diracfig}
\end{figure}

\begin{figure}
	\includegraphics[scale=0.15]{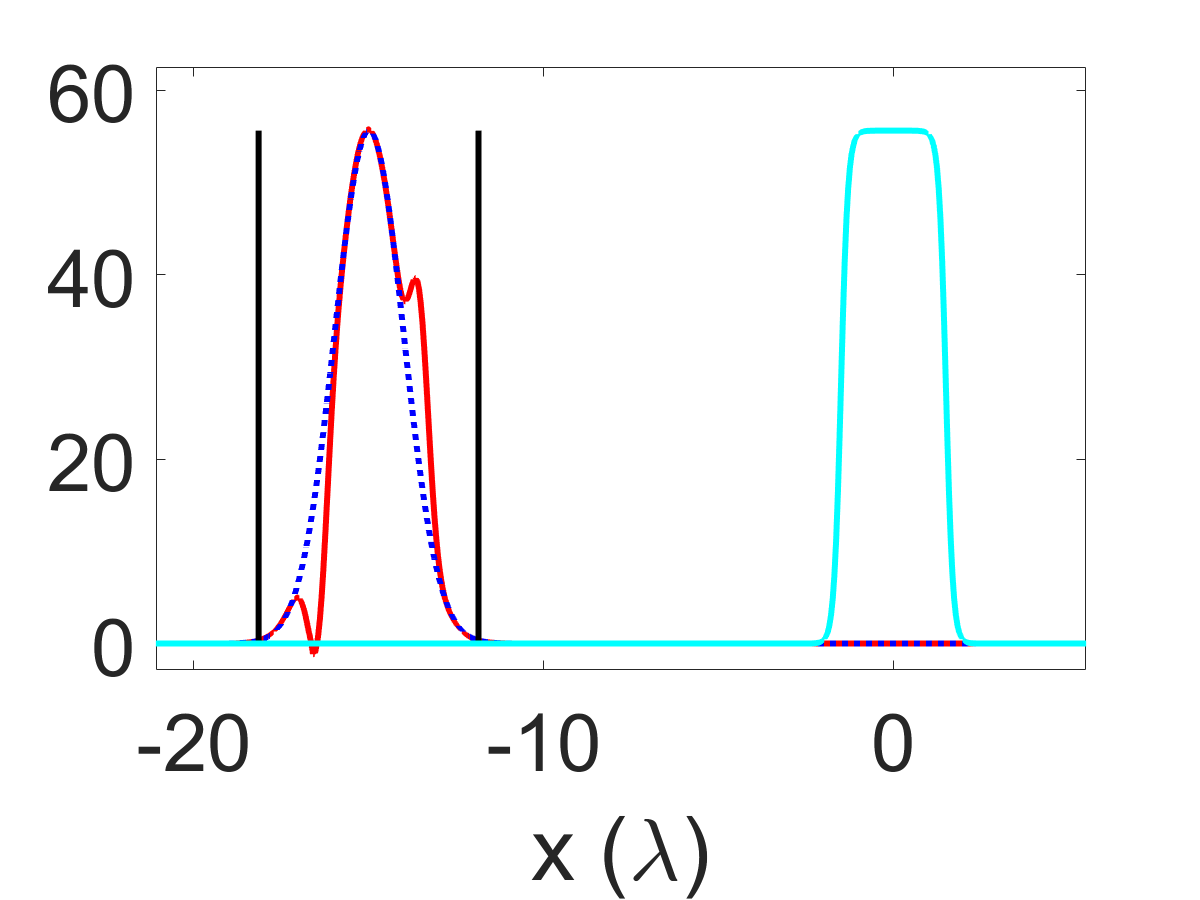}
	\includegraphics[scale=0.15]{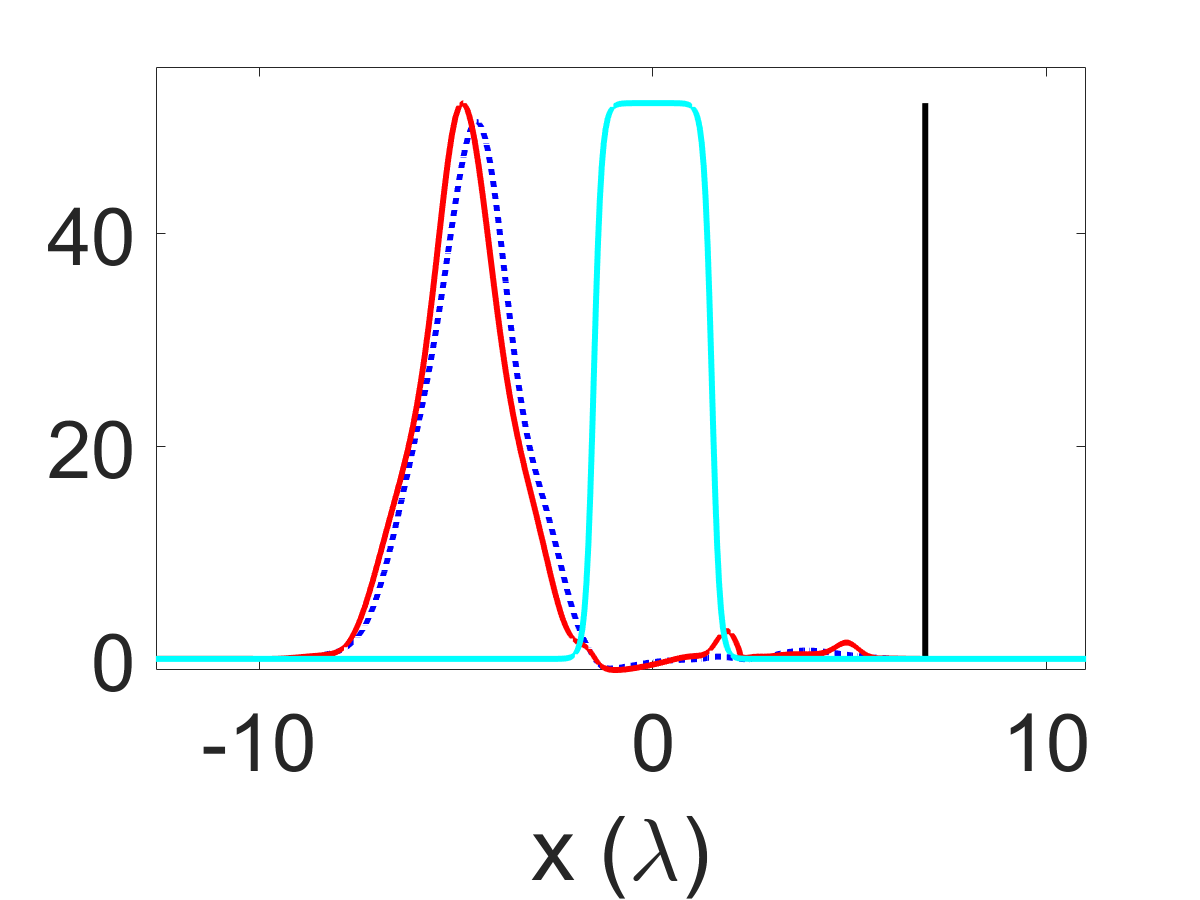}
	\includegraphics[scale=0.15]{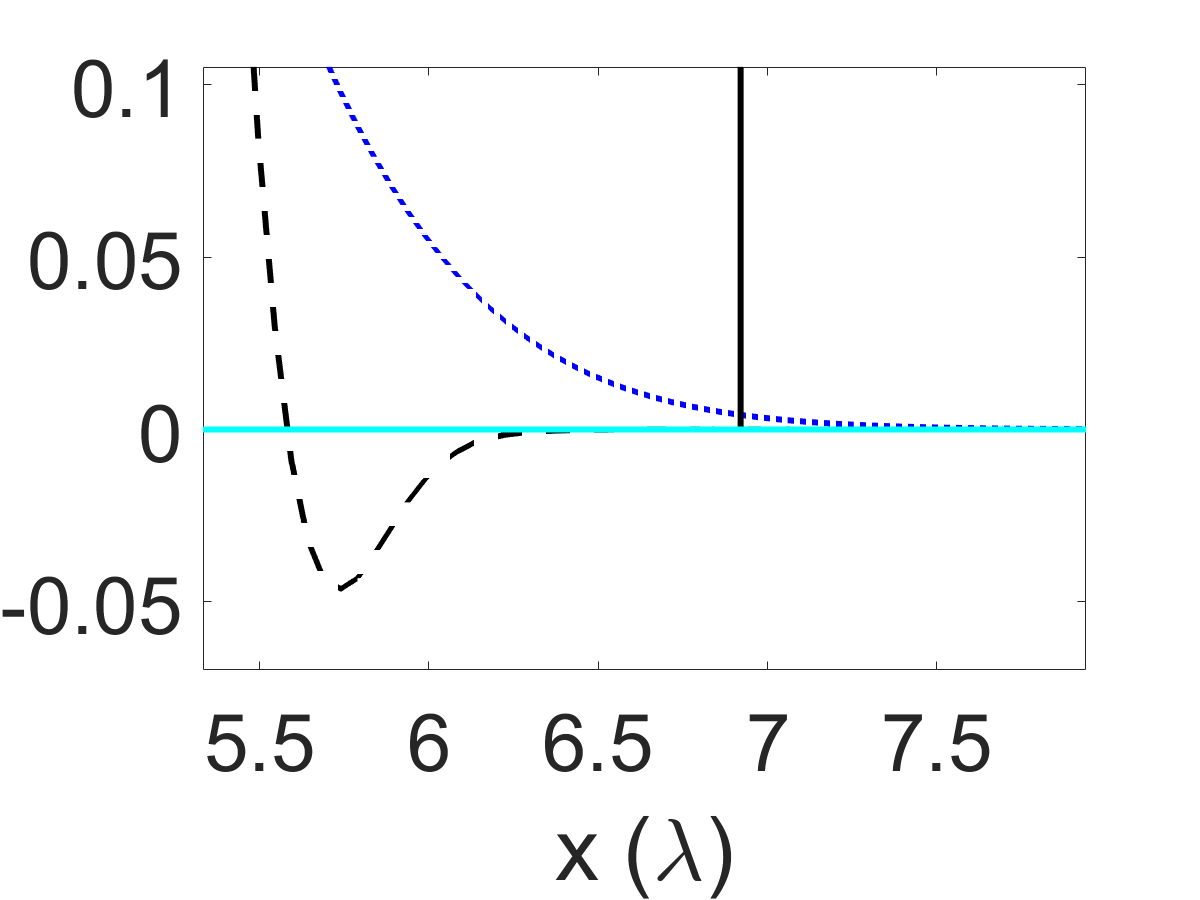}
\caption{Initial and tunneled KG wave packets, depicted using the same color scheme as in the Dirac case. The wave packet, barrier, and lattice parameters are identical to those used in the Dirac analysis, with $\lambda$ now denoting the Compton wavelength of the boson.}
	\label{kgfig}
\end{figure}

We now employ the expressions (\ref{rhoP}) and (\ref{rhoM}) in order to
compute the space-time resolved dynamics first for an initial Gaussian
wave packet, and then for that same wave packet but modified inside
$\mathcal{D}$ in the setting described above, where the "mutilation" function is
given by Eq. (\ref{mutif}) and the background potential by Eq. (\ref{poti}).
We give an illustration for a Dirac fermion in Figure~\ref{diracfig}, and another for a scalar boson described by the Klein-Gordon equation in Figure~\ref{kgfig}. The details of the computational method are given elsewhere \cite{AMpra22}.

In both cases, we can observe that the Gaussian wave packet and the mutilated one coincide outside the light cone, while they differ markedly inside the light cone (see the dashed lines in Figures~\ref{diracfig},~\ref{kgfig}). This illustrates that the intervention introduced into the initial wave packet has propagated subluminally, with no observable effect outside the light cone.

\section{Discussion and Conclusion}

\label{sec-conc}

We have seen that microcausality implies that the tunneling dynamics must remain causal,
precluding superluminal behavior. More specifically, we have proved that 
an intervention on the initial wave packet density does not modify the density outside the light cone emanating
from the region over which the intervention was performed. As a corollary, we obtained that if
two initial wave packets have a different density inside a region $\mathcal{D}$ having compact support, the density 
at a space-like location from $\mathcal{D}$ is identical for both initial wave packets. We have illustrated these results by carrying out numerical computations of space-time resolved densities for two initial wave packets differing in a region
centered on their maximum. 

While claims of superluminal or even  instantaneous wave packet 
tunneling times are common in the recent literature, both in experimental and theoretical works,
be it within a nonrelativistic or a relativistic framework (see Refs cited in \cite{causal-tunneling-PRA2025}), such
claims conflict with microcausality, a cornerstone of relativistic quantum field theories. 
Although in some cases identifying a culprit might seem straightforward  (in particular many experimental observations rely on nonrelativistic models to set a time-tag), in other cases reconciling causal dynamics with the superluminal behavior of certain quantities (such as the conditional expectation values taken on the transmitted wave packet, or on some instances of group velocities) still needs to be clarified. It would also be interesting to connect our present results to other approaches to QFT tunneling like the use of effective actions or perturbative expansions in an external potential \cite{kim,zelinski} or tunneling in the context of false vacuum decay \cite{vacuum-rev}.

\bigskip\textit{Acknowledgments.}
M. Alkhateeb acknowledges support from the C2W (Come to Wallonia) COFUND fellowship, funded by the European Union's Horizon 2020 research and innovation programme under the Marie Sk{\l}odowska-Curie
 grant agreement No. 101034383.


\bigskip



\end{document}